# Low-frequency Phonon at Perovskite Oxide Interface Studied by Surface-specific Nonlinear Terahertz Spectroscopy

Jiaming Le[1,†], Yudan Su[1,2,†], Junying Ma[1,†], Long Cheng[3], Y. Ron Shen[1,2], Xiaofang Zhai[3], Chuanshan Tian[1,4,*]

[1]Department of Physics, State Key Laboratory of Surface Physics and Key Laboratory of Micro- and Nano-Photonic Structure (MOE), Fudan University, Shanghai 200433, China

[2]Department of Physics, University of California, Berkeley, California 94720, USA

[3]School of Physical Science and Technology, ShanghaiTech University, Shanghai 201210, China

[4]Collaborative Innovation Center of Advanced Microstructures, Nanjing, 210093, China

† These authors contributed equally to this work.

*Correspondence to: cstian@fudan.edu.cn (C.T.).

**The low-frequency collective excitations, which often occur in the terahertz or multi-terahertz spectral region, play an essential role in many novel emergent phenomena. Despite numerous studies in the bulk, detection of such excitations at interfaces remains challenging owing to the lack of feasible experimental techniques. Here, we show that interfacial low-frequency modes can be characterized using surface-specific nonlinear terahertz spectroscopy. This technique uses intra-pulse difference frequency mixing (DFM) process that can extend the second-order optical spectroscopy to the terahertz range. As a demonstration, the surface phonon of $SrTiO_3$(001) at 2.8 THz was successfully measured. This surface polarization originates from the excess of oxygen vacancies or charge transfer at the interface. We have also developed an analytical procedure for remote measurement of the interfacial potential of complex oxides in a practical environment. Our method offers new opportunities for *in situ* studies of the low-frequency excitations at interfaces in broad disciplines.**

The surface and interface of complex oxides attract enormous research attention due to their unique electrical, magnetic, and electrochemical properties[1,2]. Among these oxides, strontium titanate [$SrTiO_3$ (STO)], as a prototypical perovskite retaining its multifunctional nature and the uniqueness in fabrication and modification with atomic-level precision, stands out as an ideal test-bed for exploring multifarious intriguing physical and chemical phenomena[3], in which the collective excitation and coupling present at surface/interface play the key role[4,5]. For example, at FeSe/STO interface, it is believed that both low-frequency phonon from STO and interfacial band bending are vital for the enhancement of superconductivity[6,7]. The nontrivial topological vortex/antivortex forming at the interface of $PbTiO_3$/$SrTiO_3$, the collective resonance of which lies in the terahertz (THz) range, provides an alternative choice for post-Moore electronic devices[8]. In the case of STO-based photocatalysis, the facet-dependent surface potential of STO is considered as the essential factor for electron-hole separation and charge transfer across the interface, through which water splitting with almost unity quantum efficiency can be realized[9-11]. Despite the numerous emergent interfacial phenomena explored in the past decades, the interrogation of the fundamental excitation present at surface/interface in the THz range remains experimentally challenging due to the lack of appropriate tools, especially for complex oxides surface/interface under non-vacuum or buried condition. In this regard, second-order nonlinear optical spectroscopy possesses unique advantages.

Second-order nonlinear optical spectroscopy, such as sum-frequency spectroscopy (SFS), has found multidisciplinary applications in the study of surfaces and interfaces. Being an all-optical detection scheme, it can be adopted to probe electronic or vibrational resonances in various complex interfacial systems with sub-monolayer sensitivity[12]. However, for resonant frequency below 15 THz, employing SFS is extremely difficult because of lacking intense THz light source and a feasible detection scheme that is able to distinguish the weak sum-frequency signal from the pump light. As a result, over the past decades, SFS is limited to the system with resonant frequencies ranging from mid-infrared to ultraviolet. On the other hand, for complex oxides, such as STO, many fundamental excitations related to the afore-mentioned emerging phenomena lie in the THz range. Thus, there is an urgent need for a novel surface-specific spectroscopic technique operating in the THz range. Although several attempts to observe THz response from the interface of Van de Waals layered materials have been made with the help of electronic resonances[13], a versatile THz spectroscopic scheme for probing the surface or buried interface of a complex oxide is not yet available. Besides complex oxides, such a novel technique is expected to benefit the studies of chiral vibrations of biomolecules and collective motions of hydrogen-bonding network at interfaces as well[14,15].

The surface-specific terahertz spectroscopy is based on difference-frequency mixing (DFM) process. Being a second-order nonlinear process, DFM is forbidden in the centro-symmetric bulk medium under electric dipole approximation, but is allowed at the surface/interface, where translational continuity is necessarily broken. As for THz difference-frequency spectroscopy (THz-DFS), akin to the formalism of SFS, the field spectrum of generated THz pulse can be expressed as[12]:

$$E_{THz}(\Omega) \propto \chi^{(2)}_{s,eff}(\Omega) \int d\omega \times E(\omega) E^*(\omega-\Omega) \quad (1)$$

with

$$\chi^{(2)}_{s,eff}(\Omega) = \chi^{(2)}_s(\Omega) + \chi^{(3)}_B(\Omega) \int E_{dc}(z) e^{i\Delta k z} dz \quad (2)$$

and $E(\omega)$ is the spectrum of incident pump pulse. Here, $\chi^{(2)}_{s,eff}(\Omega)$ contains both $\chi^{(2)}_s(\Omega)$ from surface structure and $\chi^{(3)}_B(\Omega) \int E_{dc}(z) e^{i\Delta k z} dz$ from surface-field-induced polarization in the depletion layer with $\chi^{(3)}_B(\Omega)$ being the third order nonlinear susceptibility of the bulk[16]. $\Delta k$ is the phase mismatch along surface normal. When $\Omega$ approaches the resonant frequency of $\chi^{(2)}_s(\Omega)$ and/or $\chi^{(3)}_B(\Omega)$, the DF signal $\chi^{(2)}_{s,eff}$ will be enhanced. If coherence length $l_c = 1/\Delta k$ is much larger than the thickness of the depletion layer, the second term on the righthand side of Eq. (2) reduces to

$$\chi^{(3)}_B(\Omega) \int E(z) e^{i\Delta k z} dz \cong \chi^{(3)}_B \int E(z) dz = \chi^{(3)}_B \Phi \quad (3)$$

where $\Phi$ symbolizes the surface potential[17]. Thus, $\chi^{(2)}_{s,eff}$ can be used to characterize both the surface structure and the surface potential $\Phi$. In the study of a buried interface or multilayers, the main advantage of THz-DFS over second harmonic spectroscopy is that the former probes a selected side of target interface with chemical sensitivity, while the latter contains contributions from all interfaces[18].

Experimentally, the THz radiation is generated via intra-pulse DFM of a femtosecond pump pulse at a surface/interface (Fig. 1a). Generally, the magnitude of $\chi_{s,eff}^{(2)}$ for a typical surface/interface is in the range of $10^{-20} \sim 10^{-22}$ m$^2$/V. For 1-TW/cm$^2$ pump intensity, the field strength of THz output is estimated at the order of $10^0 \sim 10^{-2}$ V/cm. Such a terahertz output can be recorded using state-of-the-art electro-optic sampling (EOS)[19]. To verify the practicability of THz-DFS for probing the low-frequency resonances at surfaces and interfaces, STO surface was chosen as the representative. The schematics in Fig. 1b shows the key components of THz-DFS measurement apparatus (see methods for details). A typical THz-DFS waveform from a pristine undoped STO(001) surface measured in dry air is presented in Fig. 2a, with *p*-, *p*- and *p*- (*ppp*) polarization combination for the THz and the femtosecond pump fields. After Fourier transform, a resonant feature can be clearly recognized as shown in the inset. The linear dependency of THz amplitude on pump intensity confirms the signal resulting from the second order nonlinear process (see discussion in sec. 1 of supplementary information, SI).

Since the structure of STO is centro-symmetric, the second order nonlinear susceptibility vanishes in the bulk. The surface polarization of STO was proposed to originate from the structure of the topmost layer and/or the surface-electric-field-induced polarization in the depletion layer[20]. In the former case, the resonant frequency and linewidth of the surface mode ($\chi_s^{(2)}(\Omega)$) is expected to differ from those in the bulk because of the uniqueness of lattice structure at surface/interface versus its bulk counterpart. In the latter, the lattice structure in the depletion layer is polarized within the length of screening when a band bending occurs in the interfacial region. Thus, the characteristics of resonant mode of $\chi_B^{(3)}(\Omega)$ resembles to those in the bulk and are insensitive to surface modification. After removing the Fresnel factors (see sec. 2 of SI), the resultant amplitude, imaginary and real parts of $\chi_{s,\text{eff}}^{(2)}$ spectrum of STO is given in Fig. 2b, c and d, respectively. The spectrum presented was normalized against a z-cut α-quartz crystal. A single resonance peak is recognized at 2.8 THz with 0.7-THz bandwidth (FWHM). As compared in Fig. 2e, the resonant features of $\chi_{s,\text{eff}}^{(2)}$ of STO share the same resonant frequency and linewidth with the TO1 phonon of bulk STO[21]. Besides, we modified the STO surface via deposition of a 20-nm-thick $Al_2O_3$ film. No change in spectral feature was observed except for the increase of amplitude. These results signify that the measured $\chi_{s,\text{eff}}^{(2)}$ spectra of both air/STO and $Al_2O_3$/STO interfaces are dominated by the $\chi_B^{(3)}(\Omega)$ term in Eq. (2). The variance of their amplitude is attributed to the difference of their surface potential $\Phi$.

The surface potential of STO can ascribe to surface oxygen vacancy[22,23] and/or charge transfer at its heterogeneous interface that causes band bending in the interfacial region[24]. Notably, following Eq. (2), if the $\chi_B^{(3)}$ spectrum can be quantitatively determined, one can deduce the potential of STO interface in various complex environments, e.g., in ambient/liquid environment or buried under hetero-interfaces, following Eq. (2). In order to calibrate $\chi_B^{(3)}$ of STO, we coated titanium films on both sides of a 0.5-mm-thick STO(001) plate as electrodes to apply a DC voltage, as sketched in the inset of Fig. 3a and then measured its THz-DF spectrum at different voltages. For such a metal/semiconductor interface, $E(z)$ can be quantitatively described with the well-known Schottky barrier model[25],

$$E(z) = \begin{cases} qN_d(D-z)/\varepsilon_0\varepsilon_r, & 0 < z < D; \\ 0, & z > D. \end{cases} \quad (4)$$

Here, $N_d$ is carrier concentration in bulk, $D$ is the depletion layer thickness of STO surface defined by $D = [2\varepsilon_0\varepsilon_r\Phi/eN_d]^{1/2}$ and $\varepsilon_r$ is the dielectric constant of STO. $\Phi$ is the potential drop across the Ti/STO interface. Combined with Eq. (2), the quantity of $\chi_B^{(3)}$ and $N_d$ can be deduced by fitting a set of $\chi_{s,eff}^{(2)}$ spectra with various applied voltage to the measured data (see Sec. 3 in SI), where the surface contribution ($\chi_s^{(2)}(\Omega)$) is smaller than $\chi_B^{(3)}\Phi$ by orders of magnitude. Fig. 3a and b show THz-DFS waveforms and the corresponding $\chi_{s,eff}^{(2)}$ spectra with applied voltage ramping from +1 to +200 V, the sign of which is defined as depicted in the inset of Fig. 3a. The spectral amplitude increases monotonically with applied voltage and the voltage dependent $\chi_{s,eff}^{(2)}$ curve was fit well with the above model as plotted in Fig. 3b. From fitting we acquired the third-order susceptibility of STO at TO1 vibrational frequency $\chi_{TO1}^{(3)}$ to be $5.8\times10^{-16}$ m$^2$/V$^2$ and the bulk carrier concentration $N_d = 8.9\times10^{-17}$ cm$^{-3}$ (Details are given in sec. 3 in SI). The latter agrees nicely with those reported in literature for an undoped STO[26-29].

With known $\chi_B^{(3)}$, we then applied the THz-DFS to investigate the potential of STO interfaces in a non-vacuum environment and buried in heterogeneous structure, which are challenging for traditional surface analyzing techniques. Size of the potential can be calculated using Eq. (3) with measured amplitude of the TO1 mode, and its sign is directly related to the phase of $\chi_{s,eff}^{(2)}$, referring to the waveform in Fig. 3a where the potential is positive. As shown in Fig. 4a and b, the amplitude of TO1 mode in $\chi^{(2)}_{s,\,eff}$ spectrum is 40% greater when STO is placed in atmosphere of pure nitrogen rather than in dry air (both at ambient pressure). Because oxygen molecules adsorbed on STO surface passivate the surface oxygen vacancies[30,31], that difference can be attributed to the oxygen in air. To confirm this hypothesis, we conducted THz-DFS measurement of STO under various oxygen pressure. The amplitude of TO1 mode and the corresponding surface potential increases with the lower oxygen pressure as depicted in Fig. 4g. In particular, at oxygen pressure of 1 Torr, the potential reaches +0.26 V, close to that measured by XPS in vacuum (+0.3 V)[32]. This result shows that the smaller surface potential of STO in ambient air is caused by oxygen adsorption. Considering STO surface as a promising and robust platform for photocatalytic reactions, THz-DFS offers a unique probe for *in situ* monitor of surface potential in a practical chemical environment.

STO is also frequently employed for discovery of emergent physical phenomena at its heterogeneous interface. Here we applied THz-DFS to the model system LaAlO/SrTiO$_3$ (LAO/STO), which is known as the host of two-dimensional electron gas (2DEG) appeared at heterointerface, and the interfacial superconductivity[33]. For 2DEG at LAO/STO interface, the veiled mechanism is still under lively debate, where the potential at LAO/STO interface is one of the key parameters that govern the electronic properties. To measure the potential at such buried interface, X-ray photo-emission spectroscopy (XPS) and electron energy loss spectroscopy (EELS) were two available options. However, their applicability was limited to ultra-thin films in vacuum because of signal loss of the electrons scattered by the residual gas molecules. Moreover, analysis of the experimental result is usually affected by the existence of adjacent surfaces and interfaces. Being an all-optical probing scheme with chemical sensitivity,

THz-DFS does not suffer from the above limitations. We measured two LAO/STO samples with thickness of LAO layer being 2 unit cell (u.c.) and 6 u.c. in dry air, respectively. It's worth noticing that 2DEG emerges when the LAO overlayer reaches the critical thickness of 4 u.c.[34]. Interestingly, the surface spectrum of STO remains the same regardless of the thickness of LAO layer, as shown in Fig. 4e and h. Since the spectrum of LAO/STO samples exhibits identical features to that of bare STO, analysis on interfacial potential is not interfered by the LAO overlayer. The interfacial potential of LAO/STO was found to be +0.45 V (see Fig. 4h), which agrees with that measured by XPS [34]. Another limitation for XPS or EELS is the thickness of the overlayer, because electrons can only escape from within a layer of a few nanometers or less beneath the surface. In contrast, THz photons can access an interface buried much deeper even if the overlayer is absorptive. Fig. 4e presents THz-DF spectrum of STO with 20-nm-thick $Al_2O_3$ overlayer. The interfacial potential with thick $Al_2O_3$ overlayer is +0.58 V, which is the same as STO with thinner (2 u.c.) $Al_2O_3$ capping layer[35]. Again, it manifests that the interfacial potential is mainly determined by the first few atomic layers of oxide at the interface.

STO doped with transition element, e.g., Nb, is widely used as a conductive substrate. How the dopants affect the interfacial properties of STO is still an open question, which can be readily addressed using THz-DFS. Here, we compared the THz-DF spectra of pristine STO(001) and Nb doped STO(001) (0.5 wt%, $N_d = 3\times10^{19}$ cm$^{-3}$) in dry air, as shown in Fig. 4a and f. In the latter spectrum, besides the TO1 phonon resonance, the Im $\chi^{(2)}_{s,\,eff}$ spectrum comprises a broad continuum that diverges towards the low frequency, which is a typical spectral feature of the response from free carriers[36]. Such response, $\chi^{(3)}_{free\,carrier} E E^* E_{dc}$, originates from the spatial gradient of carrier concentration set up in the depletion layer of Nb:STO[37,38]. Akin to Eq. (3), we use $\chi^{(3)}_{\text{free carrier}}\Phi\,(=\chi^{(3)}_{\text{free carrier}}\int E_{\text{dc}}(z)e^{i\Delta k z}\mathrm{d}z)$ to represent the free carrier contribution to the measured THz spectrum, where $\chi^{(3)}_{\text{free carrier}}$ denotes the third-order nonlinear susceptibility of the free carriers excited by two optical fields and one DC field. Thus, the total effective susceptibility of Nb:STO can be written as $\chi^{(2)}_{\text{s, eff}}=(\chi^{(3)}_{\text{TO1}}+\chi^{(3)}_{\text{free carrier}})\Phi$. The nonlinear optical response from the carriers can be well described by the hydrodynamic model of charge fluid[39]. We fit the measured spectrum given in Fig. 4f using a Lorentzian mode for TO1 phonon and a continuum line-shape for the carriers (see Sec. 4 in SI for details). The amplitude of TO1 mode obtained from fitting is found to be the same for the Nb: STO and the pristine STO, signifying that doping of Nb to the level of 0.5 wt% has negligible effect on surface potential as compared in Fig. 4h. The $\chi^{(3)}_{\text{free carrier}}$ spectrum deduced from Fig. 4f is plotted in Fig. S9b in SI. Since $\chi^{(3)}_{\text{free carrier}}$ is proportional to the carrier concentration ($N_d$)[39], the THz-DF response from STO with different doping level can be easily calculated by multiplying a factor of $N_d/3\times10^{19}$ cm$^{-3}$. Therefore, the THz-DF spectrum from free carriers can also be used to determine the surface potential of STO or other oxides. Moreover, the basic properties of carriers within the depletion layer, e.g., scattering rate, are accessible by quantitative analysis of the $\chi^{(3)}_{\text{free carrier}}$ spectrum using the hydrodynamic model (see also Sec. 4 in SI for details).

Conclusively, our work established THz-DFS as a viable surface-specific nonlinear optical spectroscopic method for characterization of low-frequency excitations at surfaces or buried interfaces. As a demonstration, STO(001) surface was investigated and the interfacial TO1 phonon was observed that originate from the surface polarization. Meanwhile, THz-DFS was proved as a feasible tool to quantify surface potential (or surface charge density) of complex oxides, even in gaseous environment or with capping overlayers, which is challenging for conventional surface analysis technique. To our best knowledge, this is the first time that the surface nonlinear optical spectroscopy reaches THz region. In contrast, the well-established sum-frequency spectroscopy found successful applications in probing the elementary vibrations in the mid-IR region. However, limited by the spectral tunability of the pump light below 20 THz, most of the sum-frequency studies focus on the chemical and biological materials composed of light element, e.g. H, C, N, O and F. Our work includes heavy-element compounds, the bonding of which typically vibrates at lower frequency, and collective excitations lies in the THz range into the suitable subjects for surface nonlinear optical spectroscopy. Although the THz-DFS spectrum of STO was dominated by the phonon in the depletion layer, detection of the topmost layer is feasible with EOS sensitivity of 0.01 V/cm[19]. Our approach opens up new opportunities for exploring the low-frequency vibrations and collective excitations on surfaces or at buried interfaces in a hostile environment. As an outlook, the detection bandwidth of THz-DFS that we demonstrated here is limited by the EO crystal. Using a high quality organic EO crystal, the detection bandwidth can reach beyond 5 THz. Research in this frequency range is intriguing because collective excitations at the interface of various condensed matter system occur between 5-15 THz, such as superconducting gap, heavy Fermion plasmons, soft mode in ferroelectricity, etc.

**METHODS**

**Experimental setup**

The experimental setup of THz-DFS is shown schematically in Fig. 1b. The system is based on a Yb:KGW regenerative amplifier (Light Conversion PHAROS) operating at 100 kHz repetition rate. The amplifier output centered at 1030 nm undergoes a nonlinear spectral broadening stage as described elsewhere[40]. After reflecting on a set of chirped mirrors, the pulse was compressed to 26 femtosecond. A tiny portion was picked out by a beam-splitter to be used as the probe pulse in EOS, while the majority pumps intra-pulse DFM process on the sample surface/interface. The incident pump was 44 μJ in energy per pulse and is focused to 0.5 mm ($1/e^2$ diameter) on sample. An achromatic half wave plate is used to rotate the pump polarization. The emitted terahertz pulse in reflecting direction was collimated and re-focused onto a 0.3-mm-thick GaP(110) crystal by a pair of parabolic mirrors with 200 mm and 100 mm focal length, respectively. Between the two parabolic mirrors, a broadband wire grid polarizer was used as the analyzer for the terahertz radiation. The reflected residual pump was filtered out by a PTFE plate. The whole terahertz beam path was purged with dry air to avoid vapor absorption. Routed through a translational stage, the probe beam was combined collinearly with terahertz beam by a Ge wafer. The polarization change of the probe in EO crystal was measured using the balanced detection scheme consisting of a silicon-photodiode-based balanced detector (Newport Nirvana) and a lock-in amplifier.

**Sensitivity of THz-DFS setup**

In our experiment, the balanced detection has sensitivity down to $5\times10^{-7}$ rad of polarization change integrating over 100 s (10 million pulses), corresponding to 15 V/m terahertz electric field strength inside GaP EO-crystal. Detectability to such weak electro-magnetic wave corresponds to a sensitivity of $\chi^{(2)}_{s,\,eff}$ down to $1\times10^{-20}$ m$^2$/V at TO1 phonon frequency (2.8 THz). Between 1.0 and 3.9 THz, the sensitivity is better than $2\times10^{-20}$ m$^2$/V, enabling broadband measurement of the $\chi^{(2)}_{s,\,\text{eff}}$ spectrum.

**ACKNOWLEDGEMENT**

The authors would like to thank Rui Peng and Tong Zhang at Fudan University for supplying sample. C.S.T. acknowledges support from the National Natural Science Foundation of China Grants (No. 12125403 and No. 11874123) and the National Key Research and Development Program of China (No. 2021YFA1400503 and No. 2021YFA1400202).

**AUTHOR CONTRIBUTION**

C.S.T., J.M.L., and Y.D.S. conceived the experiment. J.M.L., Y.D.S., and J.Y.M. performed THz-DFS measurement and data analysis. L.C. and X.F.Z. provided sample expertise and prepared samples. The paper was written by C.S.T, J.M.L., Y.D.S., and J.Y.M., with discussion from Y.R.S. All authors discussed and revised the manuscript.

**COMPETING INTERSETS**

The authors declare no competing interests.

**Figures and caption**

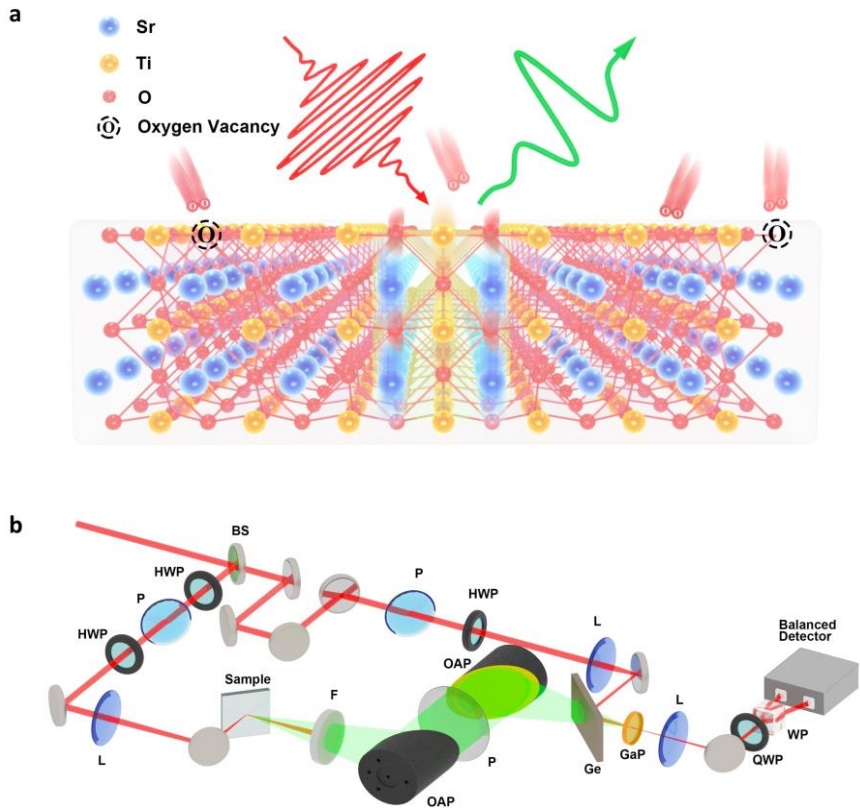

**Figure 1 | Terahertz difference frequency spectroscopy on SrTiO₃. a,** schematic of surface difference frequency generation from STO(001). The incident near-infrared pulse excites TO1 phonon vibration via intra-pulse DFM. The symmetry-breaking of TO1 phonon results from the surface field of STO that can be altered by oxygen adsorption or oxygen vacancy at the topmost layer. **b,** sketch of experiment setup. BS, beamsplitter; HWP, half-wave plate; P, linear polarizer; L, lens; F, teflon filter; OAP, off-angle parabolic mirror; QWP, quarter wave-plate; WP, Wollaston prism.

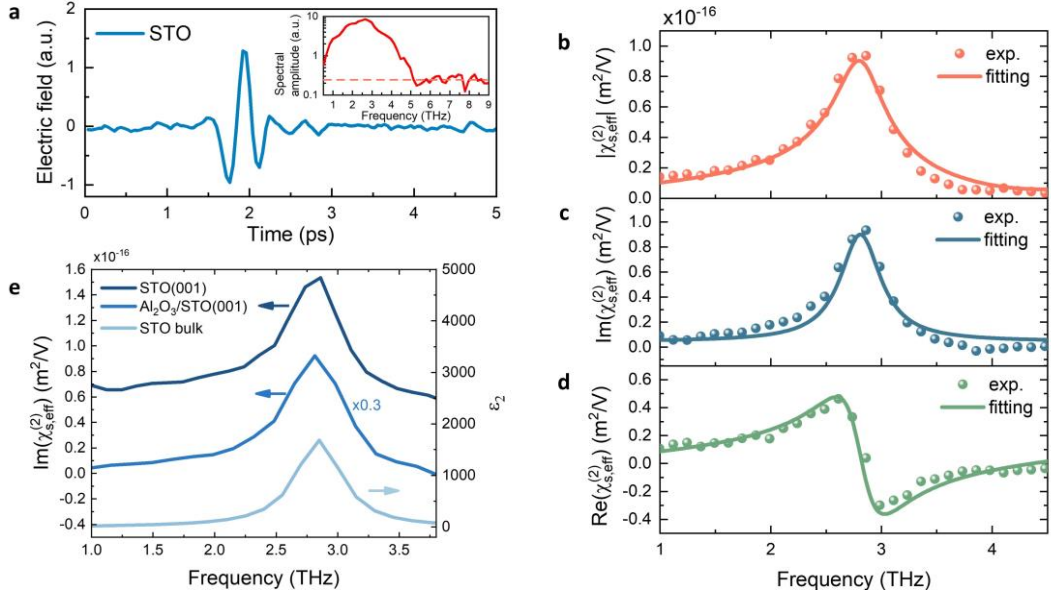

**Figure 2 | Terahertz difference frequency spectrum of SrTiO₃. a,** typical waveform of the emitted terahertz pulse from STO(001) with its Fourier transform in the inset. Dashed line in the inset represents the detection limit. **b,c,d,** amplitude (**b**), real part (**c**) and imaginary part (**d**) of deduced surface second order susceptibility, after normalization against z-cut quartz reference and removal of Fresnel factor. **e,** comparison between $\text{Im}\chi^{(2)}_{s,\text{eff}}$ of bare STO(001) (top), Al₂O₃/STO(001) (middle) and the imaginary part of dielectric function ($\varepsilon_2$) of bulk STO (bottom)[21].

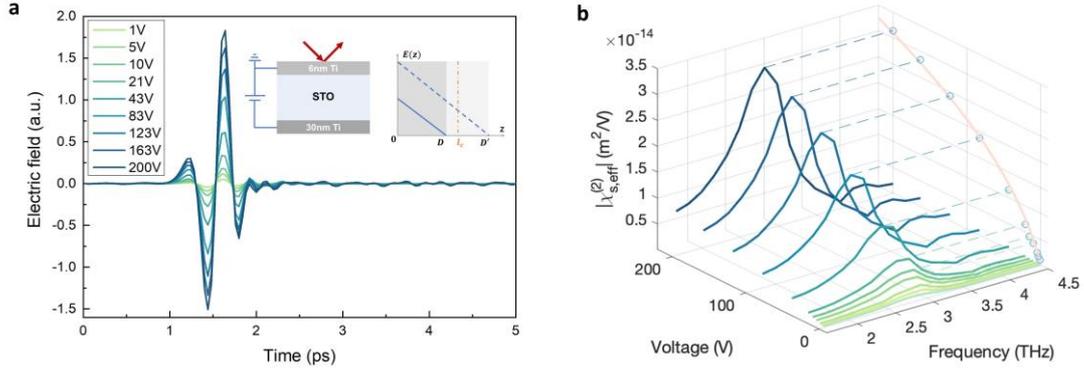

**Figure 3 | THz-DFS of STO(001) with an external applied voltage. a,** terahertz waveforms of STO with various external voltage. Inset: schematic of Ti/STO/Ti heterostructure and dependence of surface field on depth of Ti/STO Schottky junction with (dashed blue line) or without (solid blue line) external voltage. STO sample is covered with 6-nm-thick Ti film (anode) on the top and a 30-nm-thick Ti film (cathode) on the bottom. With external voltage applied, the thickness of depletion layer $D$ increases. $l_c$, coherence length of DFM process. **b,** the deduced $\left|\chi^{(2)}_{s,\text{eff}}\right|$ spectra of STO(001) under different external voltages. Blue circle, voltage-dependent amplitude of $\left|\chi^{(2)}_{s,\text{eff}}\right|$ at TO1 phonon frequency. Red curve on the right is fitted using Eq. (2) and Eq. (4) at TO1 phonon frequency.

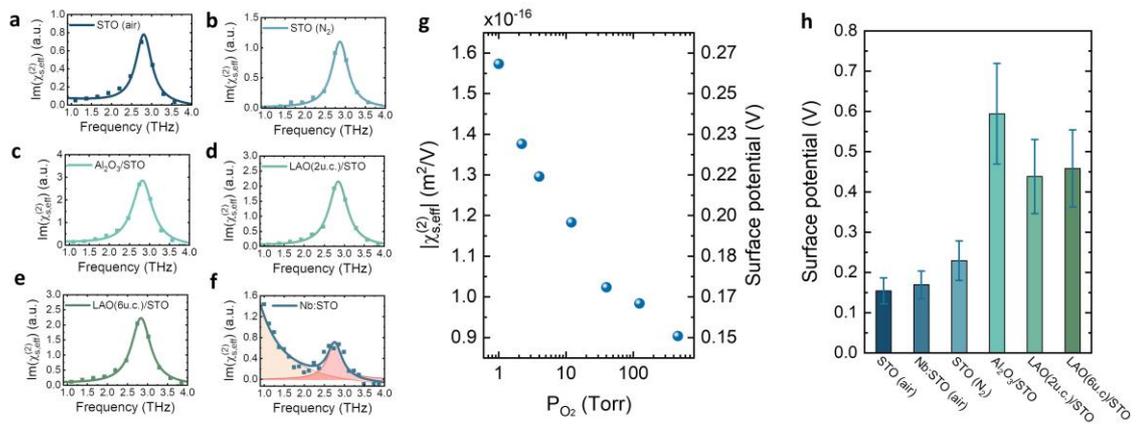

**Figure 4 | Surface/interface potential of STO in multiple conditions. a-f,** spectra of Im$\chi_{s,eff}^{(2)}$ for STO(001) in dry air (a), pure nitrogen (b), 20-nm-thick Al$_2$O$_3$/STO(001) (c), 2-unit-cell LAO/STO(001) (d), 6-unit-cell LAO/STO(001) (e), and Nb: STO(001) (f). In **f**, the fitting consists of contributions from free carriers (orange shadow) and TO1 phonon resonance (red shadow). **g,** amplitude of $\chi_{s,eff}^{(2)}$ at TO1 phonon frequency and the corresponding surface potential of bare STO(001) versus O$_2$ pressure. **h,** summarizing chart of surface/interface potential of STO(001) in various conditions.